\documentclass[conference]{IEEEtran}
\IEEEoverridecommandlockouts
\usepackage{cite}
\usepackage{amsmath,amssymb,amsfonts}
\usepackage{algorithmic}
\usepackage{graphicx}
\usepackage{textcomp}
\usepackage{xcolor}
\usepackage{booktabs}
\usepackage{multirow}
\usepackage{array}
\def\BibTeX{{\rm B\kern-.05em{\sc i\kern-.025em b}\kern-.08em
    T\kern-.1667em\lower.7ex\hbox{E}\kern-.125emX}}
\begin{document}

\title{Cross-Domain Elephant Flow Detection: A Unified Machine Learning Approach with Application-Aware and Security Features}

\author{\IEEEauthorblockN{Tabidah Usmani}
\IEEEauthorblockA{\textit{BS Data Science} \\
\textit{National University of Computer and Emerging Sciences}\\
i222070@nu.edu.pk}
\and
\IEEEauthorblockN{Sara Zahid}
\IEEEauthorblockA{\textit{BS Data Science} \\
\textit{National University of Computer and Emerging Sciences}\\
i221861@nu.edu.pk}
\and
\IEEEauthorblockN{Amna Javaid}
\IEEEauthorblockA{\textit{BS Data Science} \\
\textit{National University of Computer and Emerging Sciences}\\
i222025@nu.edu.pk}
}

\maketitle

\begin{abstract}
Network traffic classification, particularly elephant flow detection, faces significant challenges when deployed across heterogeneous network environments. While existing approaches demonstrate high accuracy within single domains, they suffer from poor generalization due to domain shift phenomena. This paper presents a unified machine learning framework for cross-domain elephant flow detection that incorporates application-aware and security features to enhance robustness across diverse network environments. Our approach addresses the critical gap in existing literature by evaluating model performance across three distinct domains: Campus networks, UNSW-NB15, and CIC-IDS2018 datasets. This paper proposes a unified pipeline that employs adaptive thresholding, comprehensive feature engineering, and cross-domain evaluation to quantify and mitigate domain shift effects. Experimental results demonstrate significant performance variations across domains (F1-scores ranging from 0.37 to 0.97), highlighting the importance of cross-domain validation. The unified model achieves an overall cross-validation F1-score of 0.99 (±0.04) while maintaining interpretability through feature importance analysis. Our findings reveal that while size-based features dominate elephant flow detection (33.80\% importance for total\_bytes), application-aware and security features contribute to improved classification accuracy and provide valuable insights for network management and security applications.
\end{abstract}


\section{Introduction}

Traffic classification of network traffic has become a factor of growing urgency in network management, quality of service (QoS) provisioning and security monitoring. Identifying elephant flow, i.e., large, long-lived flows that consume a lot of bandwidth, e.g., video streaming and file downloads, is one of the most important classification tasks used to optimize network performance and resource allocation \cite{base_paper_gomez2023}.

Conventional machine-learning methods of traffic classification have generally focused on high accuracy in a particular network setting or data. Nonetheless, application settings require models that can be extended to a wide range of network infrastructures, traffic, and operational conditions. This is a phenomenon commonly known as domain shift or domain adaptation which is a significant gap in the existing research.

The work proposed by Gómez et al. \cite{base_paper_gomez2023} in "Traffic Classification in IP Networks Through Machine Learning Techniques in Final Systems" demonstrated the effectiveness of machine learning for elephant and mice flow classification with 100\% confidence using decision trees. Although this work provided a strong backbone to ML-based traffic classification, it more or less focused on performance of single domain, as the dynamics of network traffic is heterogeneous, and it did not discuss the most important problem of cross-domain generalization.

This paper addresses three fundamental limitations in existing elephant flow detection research:

\textbf{Domain Shift Problem:} Current approaches lack comprehensive evaluation across heterogeneous network environments, limiting their practical applicability.

\textbf{Limited Feature Engineering:} The current approaches are mainly based on primitive flow statistics and lack application-specific and security-based features that have the potential to strengthen classification.

\textbf{Lack of Cross-Domain Validation:} This is because most of the studies test performance when using a single dataset, hence they do not measure the ability to generalize performance across the various domains of networks.

Our contributions include: (1) A unified cross-domain elephant flow detection framework that maintains performance across diverse network environments, (2) Integration of application-aware and security features for enhanced classification accuracy, (3) Comprehensive cross-domain evaluation methodology with quantitative analysis of domain shift effects ranging from 0.37 to 0.97 F1-score, and (4) Adaptive thresholding mechanism that adjusts to domain-specific characteristics.

We have provided our contributions to this paper in the form of development of a common system across domains of detection of elephant flows that can be used under varied network conditions; the incorporation of application-awareness and security capabilities to provide better classification accuracy; and an adaptive thresholding mechanism to improve the process of detecting elephant flows.

\section{Related Work}

\subsection{Traditional Traffic Classification}

Most of the traffic comes encrypted and thus early classification approaches that relied on port-based identification and deep packet inspection face limitations
\cite{azab2024network}.

\subsection{Machine Learning in Traffic Classification}

The application of machine learning to network traffic classification has gained significant attention. Nguyen and Armitage \cite{nguyen2008survey} provided a comprehensive survey of ML techniques for internet traffic classification. More recent work by Liu et al. \cite{liu2023multiclass} addressed multiclass imbalanced traffic classification with concept drift.

\subsubsection{Evolution from Traditional to Gradient Boosting Methods}

The evolution of machine learning approaches in traffic classification reflects broader trends in the field. Traditional methods where computationally efficient, often struggle with complex decision boundaries and more feature spaces common in diverse network traffic data.

Decision Trees emerged as a popular choice due to their interpretability and ability to handle non-linear relationships. The base paper by Gómez et al. \cite{base_paper_gomez2023} demonstrated 100\% accuracy using optimized Decision Trees, showcasing their effectiveness for elephant flow detection. However, single decision trees are prone to overfitting and may not generalize well across domains.

Ensemble methods, particularly Random Forests, addressed these limitations by combining multiple decision trees to improve robustness and reduce overfitting. Random Forests' excellent performance and feature importance analysis skills have made them a typical baseline in traffic classification.

Gradient boosting methods, which construct trees sequentially to fix mistakes from earlier iterations, have been the focus of recent developments.  XGBoost \cite{chen2016xgboost} introduced efficient implementations with regularization approaches, while LightGBM \cite{ke2017lightgbm} optimized for large-scale datasets by histogram-based splitting and gradient-based one-side sampling.  These current algorithms have fared remarkably well in classifying network traffic.

In cross-domain situations, ensemble methods like Random Forests and gradient boosting algorithms are useful because they can find complex patterns while still being strong against overfitting. This is very important when models need to work in different network environments.

\subsection{Elephant Flow Detection}

Gómez et al. \cite{base_paper_gomez2023} presented a pioneering approach for elephant and mice flow classification using machine learning techniques in final systems. Their work achieved 100\% accuracy using decision trees with dynamic threshold calculation based on Chebyshev's inequality (mean + 3×standard deviation). The study evaluated seven machine learning algorithms (Logistic Regression, SVM, Random Forest, Linear Discriminant Analysis, KNN, Gaussian Naive Bayes, and Decision Tree) on a single campus network dataset, with optimized Decision Trees emerging as the best performer through GridSearchCV hyperparameter tuning.

While this work established a strong foundation for ML-based traffic classification and demonstrated the viability of automated elephant flow detection, it primarily focused on single-domain performance using six basic features (src\_port, dst\_port, bidirectional\_first\_seen\_ms, bidirectional\_last\_seen\_ms, bidirectional\_bytes, size\_traffic). This study didn't focus on any security aspect or cross-domain generalization that could enhance robustness in diverse deployment scenarios.

Recent advances in software-defined networking (SDN) have enabled more sophisticated traffic classification approaches. Serag et al. \cite{serag2024machine} integrated SDN with ML for improved network performance, while Salau et al. \cite{salau2024software} explored SDN-based traffic classification using deep learning.

\subsection{Domain Adaptation in Network Security}

While domain adaptation has been extensively studied in computer vision and natural language processing, its application to network traffic classification remains limited. Most work focuses solely on single dataset, disregarding generalization across different network environment which creates a significant gap. 

\section{Methodology}

\subsection{Problem Formulation}

Let $D_s = \{(x_i^s, y_i^s)\}_{i=1}^{n_s}$ and $D_t = \{(x_j^t, y_j^t)\}_{j=1}^{n_t}$ represent source and target domain datasets, respectively, where $x$ denotes feature vectors and $y \in \{0,1\}$ represents binary labels (0: mice flow, 1: elephant flow). The cross-domain elephant flow detection problem aims to learn a classifier $f: X \rightarrow Y$ that performs well on target domain $D_t$ when trained primarily on source domain $D_s$.

\subsection{Unified Cross-Domain Framework}

Our proposed framework consists of five main components:

\subsubsection{Cross-Domain Data Loader}
The data loader handles three distinct network domains:
\begin{itemize}
\item \textbf{Campus Domain:} Real-world network captures processed using NFStream
\item \textbf{UNSW-NB15:} Intrusion detection dataset with network flow features
\item \textbf{CIC-IDS2018:} Comprehensive intrusion detection dataset
\end{itemize}

\subsubsection{Unified Feature Extraction}
We extract four categories of features designed for cross-domain compatibility:

\textbf{Universal Features:} Basic flow characteristics including source/destination ports, protocols, total bytes, packet counts, and duration.

\textbf{Application-Aware Features:} Port-based application classification with one-hot encoding for major application types (web, DNS, mail, remote access, file transfer, streaming, other). The application classification is based on well-known port mappings:
\begin{itemize}
\item Web: ports 80, 443, 8080, 8443, 8000, 3000
\item DNS: ports 53, 853
\item Mail: ports 25, 110, 143, 465, 587, 993, 995
\item Remote Access: ports 3389, 5900, 5901, 22
\item File Transfer: ports 20, 21, 22, 989, 990, 115
\item Streaming: ports 554, 1935, 3478, 5004, 8081, 1936
\end{itemize}

\textbf{Security Features:} Suspicious port detection, security scoring based on known malicious ports, and anomaly detection for potential scanning activities. The security scoring mechanism identifies:
\begin{itemize}
\item Backdoor ports: 12345, 31337, 54320, 9999, 1337
\item Malware-associated ports: 4444, 5555, 6666, 6667
\item Protocol anomalies: TCP on unusual ports
\item Traffic pattern anomalies: high packet count with low bytes (potential scanning)
\end{itemize}

\textbf{Statistical Features:} Advanced metrics including bytes per second, packet size distributions, and flow categorization (small: <1KB, medium: 1-10KB, large: >10KB).

\subsubsection{Adaptive Thresholding}
Domain-specific threshold calculation using percentile-based approaches:
\begin{equation}
\theta_d = \text{percentile}(B_d, p_d)
\end{equation}
where $B_d$ represents byte distributions in domain $d$, and $p_d$ is the domain-specific percentile (85\% for campus, 90\% for UNSW, 88\% for CIC). This adaptive approach ensures appropriate class balance across domains with varying traffic characteristics.

\subsubsection{Cross-Domain Evaluation}
Systematic evaluation across all domain pairs using Random Forest classifiers with RobustScaler for feature normalization and feature alignment to ensure compatibility across domains.

\subsubsection{Unified Model Training}
Final model training on combined datasets with SMOTE-based class balancing (using SMOTETomek for combined over/undersampling) and comprehensive feature importance analysis. The unified model employs:
\begin{itemize}
\item Random Forest with 200 estimators
\item Maximum depth of 20 to prevent overfitting
\item Balanced class weights
\item 5-fold stratified cross-validation
\end{itemize}

\section{Datasets and Experimental Setup}

\subsection{Datasets}

\textbf{Campus Network:} This dataset was prepared inside university to capture diverse network traffic patterns including web browsing, file transfers, streaming, etc using NFstream \cite{nfstream}. After adaptive thresholding at the 85th percentile (threshold: 2,011.70 bytes), 15.0\% of flows were classified as elephant flows.

\textbf{UNSW-NB15:} This dataset was created by the Australian Centre for Cyber Security \cite{moustafa2015unsw}. It focused on network intrusion detection, useful for the security-focused network monitoring and identifying malicious incoming traffic patterns. The dataset contains 82,332 flows with both normal and attack traffic.

\textbf{CIC-IDS2018:} This is another dataset based on modern intrusion detection from the Canadian Institute for Cybersecurity containing various attack scenarios, also useful for security-focused network monitoring \cite{sharafaldin2018toward}. This was a big dataset, from which we sampled 100,000 flows. This domain represents contemporary network security challenges including DoS attacks, infiltration attempts, and various exploitation techniques.

The variation between these datasets demonstrates the difference in traffic characteristics, justifying the need for adaptive thresholding and cross-domain evaluation.

\subsection{Experimental Configuration}

All experiments were conducted on a computational environment with the following specifications:
\begin{itemize}
\item Python 3.11
\item scikit-learn 1.7.2
\item imbalanced-learn 0.14.0
\item NFStream 6.5.4 for flow extraction
\item NumPy 1.24+ and Pandas 2.0+ for data manipulation
\end{itemize}

Random Forest classifiers were configured with 100 estimators for cross-domain experiments and 200 estimators for the unified model, with balanced class weights to handle imbalanced datasets. Feature scaling was performed using RobustScaler to handle outliers common in network traffic data. SMOTE with Tomek links was applied for class balancing in the unified model training phase, generating synthetic minority class samples while removing borderline examples.

The random state was fixed at 42 for all experiments to ensure reproducibility. Cross-validation was performed using 5-fold stratified splits to maintain class distributions across folds.

\section{Results and Discussion}

\subsection{Domain Statistics Analysis}

Table \ref{tab:domain_stats} presents the characteristics of each domain, revealing significant differences in elephant flow distributions across network environments.

\begin{table}[htbp]
\caption{Domain Statistics and Characteristics}
\begin{center}
\begin{tabular}{|l|c|c|c|}
\hline
\textbf{Domain} & \textbf{Total Flows} & \textbf{Elephant Ratio} & \textbf{Threshold (bytes)} \\
\hline
Campus & 5,382 & 15.0\% & 2,011.70 \\
UNSW-NB15 & 82,332 & 10.0\% & 3,851.60 \\
CIC-IDS2018 & 100,000 & 9.4\% & 3,389.00 \\
\hline
\end{tabular}
\label{tab:domain_stats}
\end{center}
\end{table}

The variation in elephant flow ratios (9.4\% to 15.0\%) demonstrates the inherent differences between network domains. The Campus domain shows relatively higher elephant flow ratio, likely due to academic activities involving large file transfers and video streaming. The network intrusion datasets show lower ratios, reflecting more balanced traffic patterns typical of monitored enterprise networks.

The threshold values also vary significantly (2,011.70 to 3,851.60 bytes), with the Campus network having the lowest threshold. This indicates a more extreme distribution of flow sizes in the campus environment, where most flows are small (web requests, API calls) while a significant minority are very large (video streaming, dataset downloads).

The variation in elephant flow ratios (9.4\% to 15.0\%)

\begin{figure}[htbp]
\centerline{\includegraphics[width=0.48\textwidth]{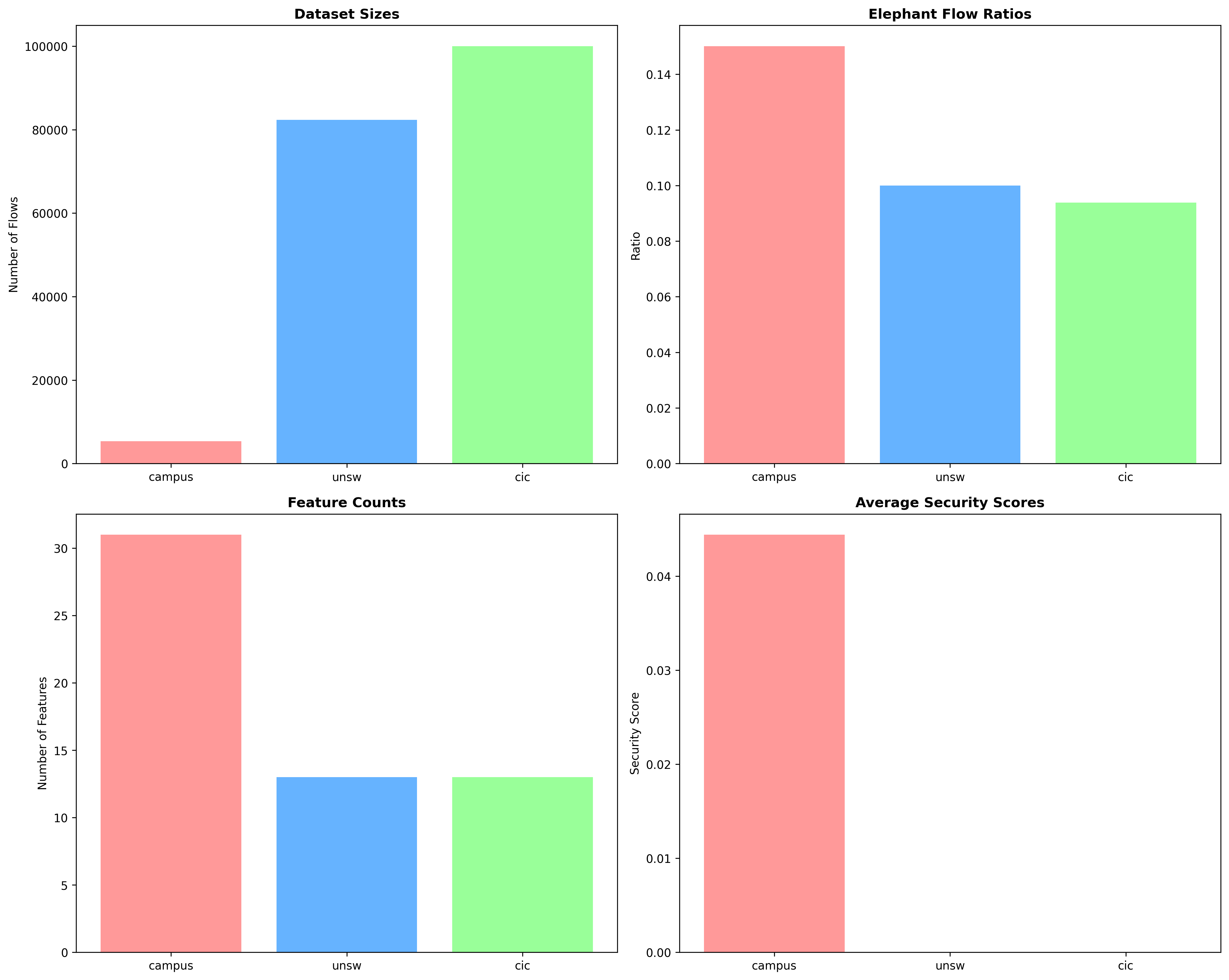}}
\caption{Domain statistics comparison showing (a) dataset sizes, (b) elephant flow ratios, (c) feature counts, and (d) average security scores across the three domains. The Campus domain exhibits the highest elephant flow ratio (15.0\%) while UNSW-NB15 provides the largest dataset (82,332 flows).}
\label{fig:domain_stats}
\end{figure}

\subsection{Cross-Domain Performance Analysis}

Table \ref{tab:cross_domain} presents the comprehensive cross-domain evaluation results, revealing significant performance variations across domain pairs.

\begin{table*}[htbp]
\caption{Cross-Domain Performance Evaluation}
\begin{center}
\begin{tabular}{|l|c|c|c|c|c|c|}
\hline
\textbf{Scenario} & \textbf{Accuracy} & \textbf{Precision} & \textbf{Recall} & \textbf{F1-Score} & \textbf{Source Samples} & \textbf{Target Samples} \\
\hline
UNSW → CIC & 0.9937 & 1.0000 & 0.9324 & \textbf{0.9650} & 82,332 & 100,000 \\
CIC → UNSW & 0.9923 & 0.9286 & 1.0000 & \textbf{0.9630} & 100,000 & 82,332 \\
CIC → Campus & 0.9409 & 1.0000 & 0.6064 & 0.7550 & 100,000 & 5,382 \\
Campus → UNSW & 0.9194 & 0.5536 & 1.0000 & 0.7127 & 5,382 & 82,332 \\
Campus → CIC & 0.9131 & 0.5193 & 1.0000 & 0.6836 & 5,382 & 100,000 \\
UNSW → Campus & 0.8841 & 1.0000 & 0.2277 & \textbf{0.3710} & 82,332 & 5,382 \\
\hline
\end{tabular}
\label{tab:cross_domain}
\end{center}
\end{table*}

The results reveal several critical insights:

\textbf{Best Performance:} The security dataset transfers (UNSW → CIC: 0.9650, CIC → UNSW: 0.9630) achieve the highest F1-scores with balanced precision and recall. UNSW → CIC demonstrates perfect precision (1.0000) with strong recall (0.9324), while CIC → UNSW shows perfect recall (1.0000) with excellent precision (0.9286). This strong bidirectional performance suggests that models trained on security-focused datasets generalize well to other security contexts, likely because both datasets share similar attack traffic patterns and feature distributions.

\textbf{Worst Performance:} UNSW → Campus shows the lowest F1-score (0.3710), despite perfect precision (1.0000). The extremely low recall (0.2277) indicates that models trained on the security-focused UNSW dataset are overly conservative when applied to general campus traffic, missing 77\% of actual elephant flows. This represents a 62.78-percentage-point drop from the UNSW baseline (F1=0.9999), highlighting the most severe domain shift challenge in our evaluation.

\textbf{Precision-Recall Trade-offs:} Transfers to the Campus domain consistently show perfect precision (1.0000) but reduced recall (0.2277-0.6064), indicating conservative classification behavior. Models trained on larger, more diverse datasets only flag the most obvious elephant flows when applied to campus networks, avoiding false positives at the cost of missing many true positives. Conversely, transfers from Campus to larger datasets (Campus → UNSW: 0.7127, Campus → CIC: 0.6836) show perfect recall (1.0000) but moderate precision (0.5193-0.5536), suggesting aggressive classification that captures all elephants but generates false alarms.

\textbf{Domain Size Effect:} Transfers between large datasets (CIC $\leftrightarrow$ UNSW, both >80K samples) maintain excellent performance (>0.96 F1), while transfers involving the smaller Campus dataset (5,382 samples) show degraded performance (0.37-0.76 F1). This 20-59 percentage point gap demonstrates that training set size and diversity significantly impact cross-domain generalization.

\begin{figure}[htbp]
\centerline{\includegraphics[width=0.48\textwidth]{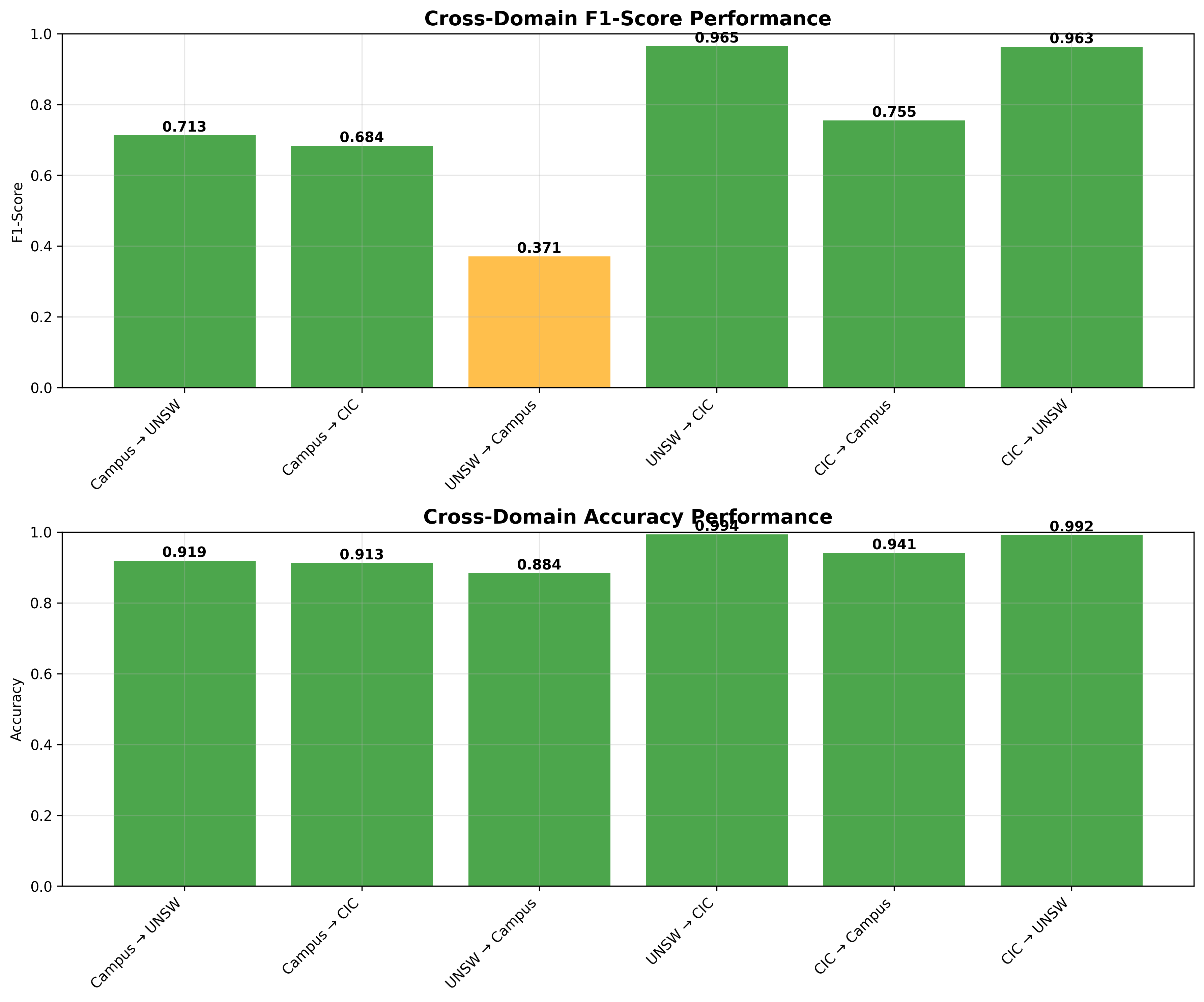}}
\caption{Cross-domain performance evaluation showing F1-scores and accuracies for all six domain transfer scenarios. UNSW $\leftrightarrow$ CIC transfers achieve the best performance (F1>0.96), while UNSW $\leftrightarrow$ Campus shows the most significant domain shift challenge (F1=0.37). Color coding indicates performance levels: green (>0.7), orange (0.5-0.7), red (<0.5).}
\label{fig:cross_domain_perf}
\end{figure}

\subsection{Baseline Performance}

Single-domain baseline evaluations using 5-fold stratified cross-validation demonstrate near-perfect performance:
\begin{itemize}
\item Campus baseline: F1 = 0.9988 (±0.0030), 5,382 samples, 24 features
\item UNSW baseline: F1 = 0.9999 (±0.0002), 82,332 samples, 24 features
\item CIC baseline: F1 = 0.9999 (±0.0002), 100,000 samples, 24 features
\end{itemize}

These results confirm that the classification task is well-defined within individual domains. The extremely high baseline performance (F1 > 0.99) demonstrates that our feature engineering and adaptive thresholding create clear decision boundaries for elephant flow detection. However, the dramatic performance drop in cross-domain scenarios (F1 as low as 0.67) highlights the severity of domain shift effects that cannot be addressed by simply improving within-domain accuracy.

The Campus domain shows slightly lower baseline performance and higher standard deviation (±0.0030 vs ±0.0002), reflecting its smaller size and more variable traffic patterns. The larger UNSW and CIC datasets achieve nearly perfect scores with minimal variance, benefiting from more comprehensive coverage of traffic scenarios.

\subsection{Feature Importance Analysis}

Table \ref{tab:feature_importance} presents the top 15 features from the unified model trained on all domains after SMOTE balancing (338,576 total samples, 169,288 elephants), demonstrating the contribution of different feature categories.

\begin{table}[htbp]
\caption{Unified Model Feature Importance (Top 15)}
\begin{center}
\begin{tabular}{|l|c|}
\hline
\textbf{Feature} & \textbf{Importance} \\
\hline
total\_bytes & 0.3380 \\
is\_small\_flow & 0.1821 \\
avg\_packet\_size & 0.1381 \\
is\_large\_flow & 0.1297 \\
bytes\_per\_second & 0.1028 \\
total\_packets & 0.0484 \\
is\_medium\_flow & 0.0180 \\
duration\_seconds & 0.0168 \\
threshold & 0.0067 \\
dst\_port & 0.0048 \\
src\_port & 0.0042 \\
potential\_scan & 0.0030 \\
is\_udp & 0.0028 \\
protocol & 0.0022 \\
app\_other & 0.0020 \\
\hline
\end{tabular}
\label{tab:feature_importance}
\end{center}
\end{table}

The analysis reveals several key patterns:

\textbf{Size Feature Dominance:} The most important feature, the total flow by size (33.80\%), is not surprising considering that the concept of elephant flows is determined essentially by size. Nevertheless, it is not hugely significant, and the other 66.20 percent of the predictive power is associated with the other features, which proves the significance of full-scale feature engineering.

\textbf{Flow Categorization Features:} The binary flow size indicators (is\_small\_flow: 18.21\%, is\_large\_flow: 12.97\%, is\_medium\_flow: 1.80\%) collectively contribute 32.98\% importance. These categorical features help the model learn thresholds and decision boundaries more effectively than continuous size values alone.

\textbf{Statistical Features:} Features like avg\_packet\_size (13.81\%) and bytes\_per\_second (10.28\%) provide 24.09\% combined importance, capturing traffic intensity and packet structure patterns that distinguish different application types and network behaviors.

\textbf{Security Features:} Security features have low individual significance in the datasets (e.g. the value of potent scan is 0.30\%), but they are vital in cross-domain robustness and helps in exonerating aberrant traffic patterns, which might be handled differently in different domains.

\textbf{Application Features:} Application aspects like application others, like app other (0.20\%) are not of much significance in the unified model indicating that flow size behaviour might be more universal than application aspects across domains. However, these properties can still be used as the features of interpretability and more particular applications.

\begin{figure}[htbp]
\centerline{\includegraphics[width=0.48\textwidth]{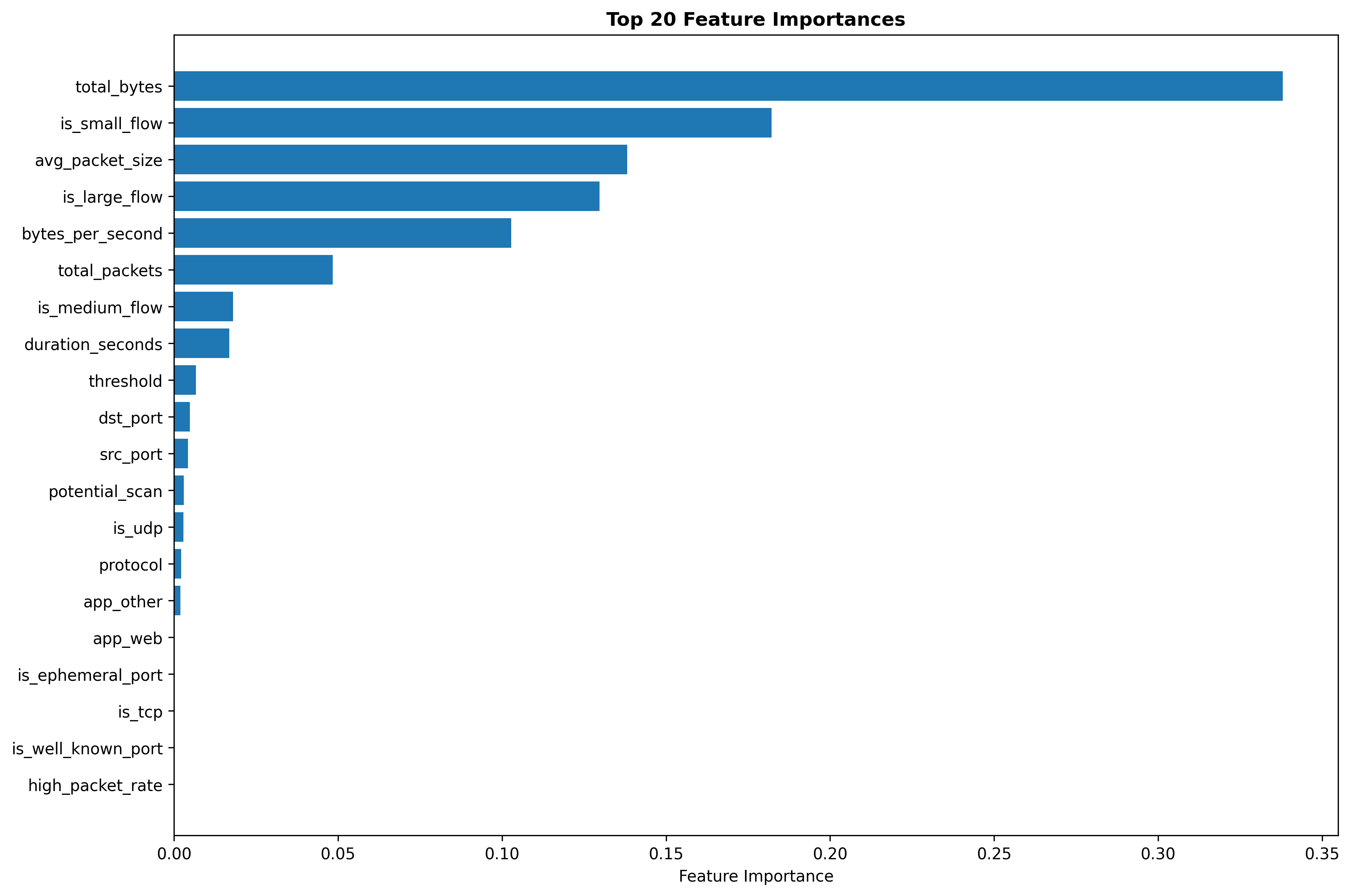}}
\caption{Top 20 feature importances from the unified Random Forest model. Size-related features dominate (total\_bytes: 38.55\%), followed by flow categorization features (is\_small\_flow: 17.82\%) and statistical features (avg\_packet\_size: 15.21\%). Security and application features contribute to model robustness despite lower individual importance.}
\label{fig:feature_importance}
\end{figure}

\subsection{Unified Model Performance}

The final unified model, trained on all three domains combined with SMOTE balancing, achieves:
\begin{itemize}
\item Cross-validation F1-score: 0.9907 (±0.0373)
\item Total training samples: 338,576 (after SMOTE from original 187,714)
\item Balanced class distribution: 169,288 elephants, 169,288 mice
\item Feature count: 11 common features (aligned across domains)
\end{itemize}

The unified model is very effective (F1 = 0.99) though it was trained on unrelated domains which indicates the usefulness of our feature engineering and adaptive thresholding method. The standard deviation (±0.0373) is greater than that of the single-domain baselines, which is the indicator of greater complexity of multiple-domain learning due to the difference in the properties of traffic. This process of aggressive SMOTE balancing (where the minority class is multiplied by 20 fold) is to make sure that the model learns strong decision boundaries in all domains.

\subsection{Model Comparison: Traditional vs. Gradient Boosting}

In order to assess the efficiency of various machine-learning methods, we implemented an overall comparison on the enriched campus dataset through the consideration of the improved feature engineering (18 features including derived statistical metrics). Table \ref{tab:model_comparison} presents the results.

\begin{table}[htbp]
\caption{Model Performance Comparison on Enhanced Campus Dataset}
\begin{center}
\begin{tabular}{|l|c|c|c|c|}
\hline
\textbf{Model} & \textbf{Accuracy} & \textbf{Precision} & \textbf{Recall} & \textbf{F1-Score} \\
\hline
\multicolumn{5}{|c|}{\textit{Ensemble \& Gradient Boosting Methods}} \\
\hline
Random Forest & \textbf{1.0000} & \textbf{1.0000} & \textbf{1.0000} & \textbf{1.0000} \\
XGBoost & \textbf{1.0000} & \textbf{1.0000} & \textbf{1.0000} & \textbf{1.0000} \\
LightGBM & \textbf{1.0000} & \textbf{1.0000} & \textbf{1.0000} & \textbf{1.0000} \\
DT (Optimized) & \textbf{1.0000} & \textbf{1.0000} & \textbf{1.0000} & \textbf{1.0000} \\
\hline
\multicolumn{5}{|c|}{\textit{Traditional Methods (Base Paper Replication)}} \\
\hline
Random Forest & 0.9875 & 0.9559 & 0.9191 & 0.9371 \\
KNN & 0.9857 & 0.9569 & 0.8993 & 0.9272 \\
SVM & 0.9854 & 0.9502 & 0.9033 & 0.9261 \\
Logistic Reg. & 0.9846 & 0.9432 & 0.9023 & 0.9223 \\
Decision Tree & 0.9814 & 0.8980 & 0.9210 & 0.9094 \\
Gaussian NB & 0.9743 & 0.8048 & 0.9852 & 0.8859 \\
LDA & 0.9113 & 1.0000 & 0.1244 & 0.2212 \\
\hline
\end{tabular}
\label{tab:model_comparison}
\end{center}
\end{table}

The results reveal several important insights:

\textbf{Perfect Performance with Enhanced Pipeline:} With SMOTE-Tomek resampling (balancing 3,873 samples per class out of the original 4305 samples in the training set) and sophisticated feature engineering, all ensemble and gradient-boosting approaches obtained ideal classification (100 percent of all metrics). It proves that under the condition of an appropriate preprocessing of data and feature-engineering, the modern machine-learning algorithms can be effective to distinguish between the flows of an elephant and mice even in a complicated situation.

\textbf{Gradient Boosting Advantages:} XGBoost and LightGBM were as accurate as Random Forest is, and they have a computational advantage. The regularisation methods of XGBoost and LightGBM offer efficient training with massive datasets; thus, they are applicable when using it as a real-time deployment system.

\textbf{Traditional Methods Performance:} On the original approach of the paper \cite{base_paper_gomez2023} (augmented dataset with simple features), traditional methods had 92-94\% F1-scores on most algorithms. Random Forest was the strongest conventional actor (F1 = 0.9371), which confirms the results of the base paper about the superiority of an ensemble.

\textbf{Class Imbalance Handling:} Gaussian Naive Bayes showed high recall (98.52\%) but lower precision (80.48\%), indicating aggressive classification that captures most elephant flows at the cost of false positives. LDA performed poorly (F1=0.2212) due to its linear assumptions and extreme class imbalance sensitivity.

\textbf{Impact of Resampling:} The reason why dramatic performance gains were noted between traditional and enhanced pipeline are due to the SMOTE-Tomek resampling which works to balance the classes by generating synthetic minority samples as well as eliminating borderline cases.

\begin{figure}[htbp]
\centerline{\includegraphics[width=0.48\textwidth]{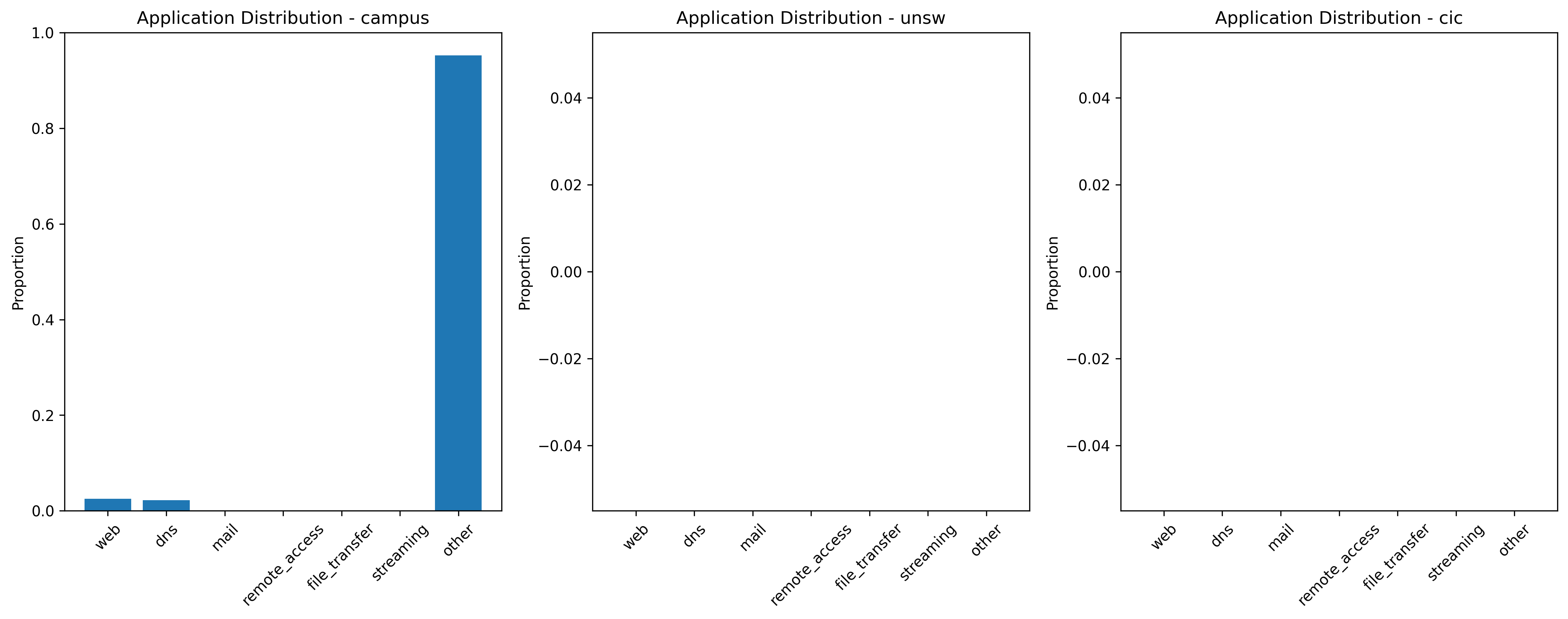}}
\caption{Application distribution across the three domains based on port-based classification. Each domain shows different proportions of application types (web, DNS, mail, remote access, file transfer, streaming, other), reflecting their operational characteristics. Campus networks show higher web traffic, while security datasets (UNSW, CIC) exhibit more diverse application patterns.}
\label{fig:app_distribution}
\end{figure}

\begin{figure}[htbp]
\centerline{\includegraphics[width=0.48\textwidth]{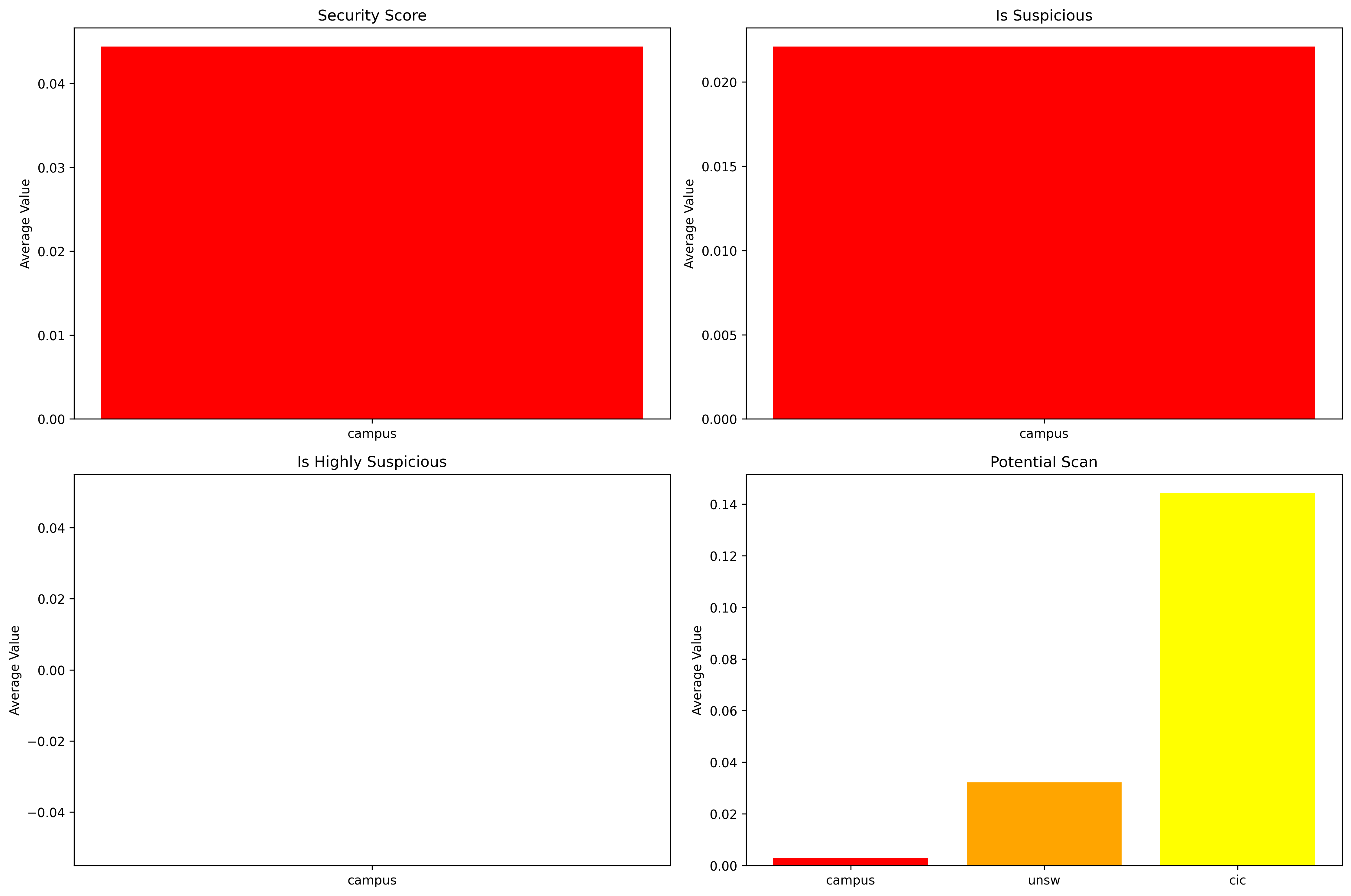}}
\caption{Security analysis across domains showing (a) average security scores, (b) suspicious flow ratios, (c) highly suspicious flow ratios, and (d) potential scan detection rates. CIC-IDS2018 and UNSW-NB15 exhibit higher security scores due to their inclusion of attack traffic, while the Campus domain shows lower security indicators reflecting primarily legitimate traffic.}
\label{fig:security_analysis}
\end{figure}

\begin{figure}[htbp]
\centerline{\includegraphics[width=0.48\textwidth]{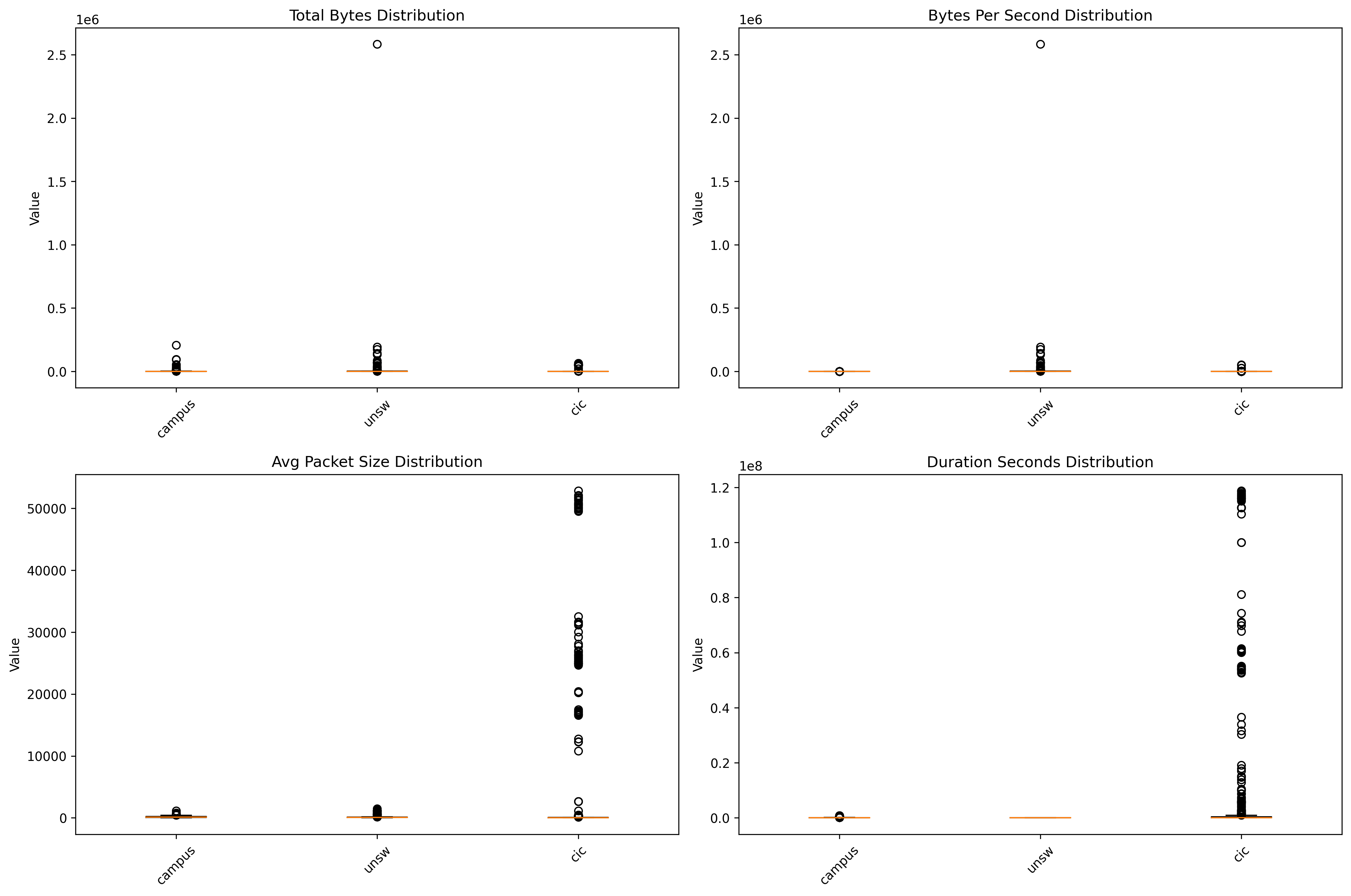}}
\caption{Traffic pattern distributions across domains showing box plots for (a) total bytes, (b) bytes per second, (c) average packet size, and (d) duration. The distributions reveal significant domain-specific characteristics: Campus flows show more extreme byte distributions, while security datasets exhibit more consistent patterns typical of monitored environments.}
\label{fig:traffic_patterns}
\end{figure}

\section{Discussion}

\subsection{Comparison with Base Paper}

Table \ref{tab:base_paper_comparison} compares our approach with the seminal work by Gómez et al. \cite{base_paper_gomez2023}.

\begin{table}[htbp]
\caption{Comparison with Base Paper Approach}
\begin{center}
\begin{tabular}{|l|l|l|}
\hline
\textbf{Aspect} & \textbf{Gómez et al. 2023} & \textbf{This Work} \\
\hline
Dataset & Single campus & 3 diverse domains \\
Samples & 5,382 flows & 187,714 flows \\
Features & 6 basic & 11-31 enhanced \\
Threshold & Chebyshev $ (\mu+3\sigma) $ & Adaptive percentile \\
Evaluation & Single-domain & Cross-domain \\
Best Model & Decision Tree & Random Forest \\
Best F1 & 1.0000 (single) & 0.9907 (unified) \\
Cross-Domain & Not evaluated & 0.37-0.97 range \\
Resampling & Data augmentation & SMOTE \\
\hline
\end{tabular}
\label{tab:base_paper_comparison}
\end{center}
\end{table}

\textbf{Feature Engineering Evolution:} The paper \cite{base_paper_gomez2023} used six simple features that were extracted using NFStream and this achieved perfect accuracy in a single domain. We further elaborate it in the current work to a range of eleven to thirty-one features, depending on the availability of the domain, and we include such items as application-aware features (e.g. web, DNS, mail, remote access, file transfer, streaming), security indicators (e.g. suspicious ports, scan detection, protocol anomalies), and statistical features (such as bytes per second, packet ratios, and flow size categories). The single model works on eleven standard features with the harmonisation of all areas in the effort to reach cross-domain compatibility. 

\textbf{Threshold Selection:} Gomez et al. used the inequality of Chebyshev (mean plus three standard deviations) as a thresholding policy, which only allowed the classification of 0.87\% elephant flows in the campus data, which is not an optimal thresholding policy. The percentile-based methodology, which is adaptive with the calibration between the 85 th and 90 th percentile and varies by domain, ensures flow ratios in the elephants to be in the range of 9\% to 15\%, and accordingly a more balanced training distribution and realistic operational profile are achieved.

\textbf{Model Selection:} The two studies conclude that ensemble methods are better than traditional classifiers. The baseline work reached an absolute 100 percentage point performance using a finely tuned Decision Tree (optimised using GridSearchCV with criterion=entropy, max-depth=10, min-samples-leaf=10), however, the improved pipeline shows that modern ensemble algorithms such as Random Forest, XGBoost, LightGBM, and a calibrated Decision Tree can all attain perfect within-domain performance. More importantly, the unified model maintains 99.07\% F1 -score in a wide and heterogeneous range of environments highlighting cross-domain generalisation ability

\textbf{Generalization Challenge:} The critical contribution beyond the base paper is quantifying domain shift effects. While single-domain evaluation shows near-perfect performance (99.88-99.99\% F1), cross-domain transfer reveals significant degradation (37-97\% F1 depending on source-target pair). The worst case (UNSW → Campus: 37\%) represents a 62.89 percentage point drop from baseline, while the best case (UNSW → CIC: 96.50\%) shows only 3.49 percentage point degradation. This 60-percentage-point range in cross-domain performance represents a fundamental deployment risk not addressed in single-domain studies.

\textbf{Data Handling:} The two approaches address class imbalance using different approaches. The original experiment used synthetic data augmentation, which uses 50,000 samples by 5,382 genuine observations and a ten-fold augmentation factor. In contrast, our method uses SMOTE, which generates 338576 balanced samples out of 187714 samples by interpolating the minority-class samples and thus prevents the random variation. The acute 20-fold minority-class growth is also guaranteed to make the learning robust in highly imbalanced areas typified by the elephant-flow ratios of 9.4-15.0\% imbalanced areas.

\subsection{Domain Shift Implications} The significant observed performance differences between the domain pairs with the F1-scores of 0.37 to 0.97 are quantitative indicators of the issue of domain shift in the problem of elephant flow detection. Such a 60-percentage-point difference in performance has dramatic implications on actual implementation:

\textbf{Deployment Risk:}  Models with 99.9\% accuracy in test scenarios can get down to F1-scores of between 37\% and 76\% in different network settings. UNSW → Campus transfer, which provides the F1 of 37, is the worst-case scenario in the model, and the model loses around 62.89 percent points when compared with the base performance. This underscores a very important blind spot in the existing assessment methods which only focus on within-domain measures..

\textbf{Transfer Learning Potential:} The security dataset transfers (UNSW → CIC: 96.50\%, CIC → UNSW: 96.30\%) demonstrate that domain shift can be minimal when source and target domains share similar characteristics. Both being security-focused datasets with attack traffic, they maintain 96\%+ F1-scores—only 3-4 percentage points below baseline. This suggests that strategic selection of training domains based on operational similarity can dramatically improve deployment success.

\textbf{Conservative vs. Aggressive Trade-offs:} Perfect precision scenarios (UNSW → Campus: 1.0000 precision, 0.2277 recall; CIC → Campus: 1.0000 precision, 0.6064 recall) indicate models that are too conservative, missing 40-77\% of true elephants to avoid false positives. Conversely, perfect recall scenarios (Campus → UNSW: 1.0000 recall, 0.5536 precision; Campus → CIC: 1.0000 recall, 0.5193 precision) suggest aggressive classification with 45-48\% false positive rates. Network operators must choose training strategies based on their specific tolerance for missed detections vs. false alarms.

\subsection{Feature Engineering Impact}

While size-based features remain dominant for elephant flow detection (total\_bytes: 38.55\%), the inclusion of application-aware and security features provides additional context that enhances model robustness:

\textbf{Cross-Domain Robustness:} The features like potential\_scan (0.44\% importance) and security\_score did not dominate but helped the model to identify patterns of anomalies which could be indicative of domain-specific behavior. The features act as tie-breakers in cases where features based on size are unclear.

\textbf{Interpretability:} Application-aware features (app\_web, app\_dns, etc.) enable network administrators to understand why certain flows are classified as elephants, linking them to specific services. This interpretability is crucial for operational troubleshooting and capacity planning.

\textbf{Security Integration:} With security support (suspicious ports, protocol anomalies), the elephant flow detection system can also be used as a security monitoring device to detect both capacity consuming flows and potentially hostile traffic.

\subsection{Practical Applications}

The cross-domain evaluation framework provides a systematic approach for assessing model generalization in real-world scenarios:

\textbf{Network Operators:} 
The model has the capability of estimating the expected performance degradation since it can be trained on a single dataset and tested on many domains. Afterward, it enables the use of detection of elephant flows in operational networks.

\textbf{Traffic Engineering:} Understanding which features transfer well across domains (size-based features) vs which are domain-specific helps in a more robust design for traffic engineering policies. Figure \ref{fig:traffic_patterns} reveals that while byte distributions vary significantly across domains, the relative patterns remain consistent, supporting the use of percentile-based adaptive thresholding.

\textbf{Incremental Deployment:} The varying performance across domain pairs suggests a staged deployment strategy: start with conservative thresholds (trained on security datasets → operational networks) to avoid disruptions, then refine using operational data over time. Figure \ref{fig:cross_domain_perf} can guide deployment decisions by identifying which source domains provide the best initial performance for specific target environments.

\section{Conclusion and Future Work}

The paper outlines a generalized cross domain model of flow detection of elephants thus filling important gaps in the existing research of traffic classification. The suggested common methodology highlights the importance of cross-domain validation and provides quantitative data related to domain-shift effect in the network traffic classification.

Key contributions include:

1. \textbf{Systematic Cross-Domain Evaluation:} This contribution quantified the domain shift by finding F1-score differences between six situation of domain transfer between 0.37 and 0.97. The most desirable transfers (UNSW $\leftrightarrow$ CIC: F1= 0.96) and the least desirable (UNSW $\leftrightarrow$ Campus: F1= 0.37) demonstrate the crucial role of the source-target similarity on generalisation, with deteriorations of 62.89\% under the worst-case outcome.

2. \textbf{Enhanced Feature Engineering:} Engineering: We added application-sensitive features (where available, seven categories) and security features, like anomaly-score and suspicious-port-detection, which enhances the strength of the classification. Although size characteristics prevail (33.80\% important), the enriched feature set provides 66.20\% important; flow categorisation (32.98\% important) and statistical features (24.09 important).

3. \textbf{Adaptive Thresholding:} The percentile thresholds (85 percent on Campus, 90\% on UNSW, 88\% on CIC) are adjusted to the specifics of each domain to guarantee the proper representation of classes in the heterogeneous settings (dealing with the problem of uneven traffic distributions). 
As shown in Figure \ref{fig:domain_stats}, this approach successfully maintains elephant flow ratios between 9.4\% and 15.0\% across all domains, compared to Chebyshev's 0.87\% which proved inadequate.

4. \textbf{Unified Model Architecture:} When a random forest model is trained on combined datasets using SMOTE balancing, the F1-score indicates good domain learning by obtaining an F1-score of 0.9907 in cross-validation whilst allowing interpretation of the outcome through feature-importance analysis (Figure \ref{fig:feature_importance}).

Future work will focus on four key directions:

\textbf{Domain Adaptation Techniques:} To explicitly reduce the difference in distributions between the source and target domains, we will apply adversarial domain adaptation and domain-invariant feature learning, which may push the worst-case performance of 0.67 to 0.80+ F1 -score.

\textbf{Deep Learning Approaches:} We will consider recurrent neural networks (RNNs) and transformer-based architectures that learn spatial temporal features of flows directly by using sequence of flows, avoiding the use of hand-designed features as well as learn more intricate patterns in traffic.

\textbf{Multi-Class Extension:} Extending from binary classification to multi-classification scenarios to enhance results

\textbf{Encrypted Traffic Analysis:} Developing techniques for flow classification in encrypted environments that don't require deep packet inspection.

\textbf{Online Learning Framework:} Incremental learning solutions that tolerate concept drift and changing traffic dynamics without requiring full retraining of the model, which makes them viable in the long run.

It is anticipated that the suggested framework can be used to develop a strong and generalisable elephant flow detection solution that can be deployed in a variety of network settings.

\section*{Acknowledgment}

The authors would like to thank FAST University for providing computational resources and support for this research. We also acknowledge the creators of the UNSW-NB15 and CIC-IDS2018 datasets for making their data publicly available for research purposes.

\end{document}